\documentclass[preprint,prl,superscriptaddress,amsmath,amssymb,floatfix,letterpaper]{revtex4}

\usepackage{graphicx}
\usepackage{dcolumn}
\usepackage{bm}
\usepackage{float}
\usepackage{amsmath}
\usepackage[latin1]{inputenc}

\begin{document}

\title{Kondo Physics in Nanotubes: \\ Magnetic-field dependence and singlet-triplet Kondo}

\author{C. H. L. Quay}

\author{John Cumings}
\altaffiliation[Present address ] {Department of Materials Science \& Engineering, University of Maryland, College Park, MD 20742-2115.}
\affiliation {Physics Department, Stanford University, Stanford, CA 94305-4060, USA.}

\author{S. J. Gamble}
\affiliation{Applied Physics Department, Stanford University, Stanford, CA 94305-4090, USA.}


\author{R. de Picciotto}
\affiliation{Bell Laboratories, Lucent Technologies Incorporated, Murray Hill, New Jersey 07974, USA.}

\author{H. Kataura}
\affiliation{Nanotechnology Research Institute, National Institute of Advanced Industrial Science and Technology, Central 4, Higashi 1-1-1, Tsukuba, Ibaraki 305-8562, Japan.}

\author{D. Goldhaber-Gordon}
\affiliation{Physics Department, Stanford University, Stanford, CA 94305-4060, USA.}


\begin{abstract}
In a single-walled carbon nanotube, we observe the spin-1/2 Kondo effect. The energy of spin-resolved Kondo peaks is proportional to magnetic field at high fields, contrary to recent reports. At lower fields, the energy falls below this linear dependence, in qualitative agreement with theoretical expectations. For even electron occupancy, we observe a spin-1 Kondo effect due to the degeneracy of the triplet ground states. Tuning gate voltage within the same Coulomb diamond drives a transition to a singlet ground state. We also independently tune the energy difference between singlet and triplet states with a magnetic field. The Zeeman splitting thus measured confirms the value of the g-factor measured from the spin-1/2 Kondo feature. \end{abstract}

\maketitle

\sloppy Carbon nanotubes, discovered almost two decades ago~\cite{ijima}, have proven to be a rich playground for physicists investigating zero- and one-dimensional quantum mechanical phenomena, such as the Kondo effect~\cite{nygard} and Luttinger-liquid behavior~\cite{bockrath:nat, yao}. The energy scales of nanotube quantum dots are particularly convenient for studying the Kondo effect: the Kondo temperature can be much larger than accessible temperatures but smaller than Zeeman splitting at accessible magnetic fields. 

In this Letter we report measurements of the low-temperature (250mK-6K) conductance of a carbon nanotube. Our device exhibits Coulomb blockade behavior and also, because of its high conductance, the Kondo effect in its spin-1/2 and spin-1 manifestations. For spin-1/2 Kondo, the energy of spin-resolved resonances falls below the Zeeman energy at low fields, but saturates to the Zeeman energy more rapidly than expected with increasing field~\cite{costi:chap,moore}. At high fields, we obtain the most precise measure to date in a quantum dot of both the g-factor and the extrapolated zero field splitting of Kondo resonances. In the case of spin-1 Kondo, we observe a gate-induced transition between triplet and singlet ground states. Singlet-triplet Kondo has been observed in GaAs quantum dots, where the transition was driven by a magnetic field coupling to orbital states~\cite{sasaki} or by a
gate voltage~\cite{kogan:prb}. In our system, because of the larger g-factor and smaller dot area, we are able to use the Zeeman effect to cleanly and independently tune through the transition, with no noticeable orbital effect of the magnetic field. This measurement provides additional confirmation of the g-factor, and is the first such quantitative investigation of the singlet-triplet Kondo effect.

This work grew out of a study of an ensemble of C$_{60}$ peapods: carbon nanotubes filled with buckyballs. Device fabrication was standard for nanotube transport studies except that we attempted to fill the tubes with C$_{60}$ molecules~\cite{fabrication}. A detailed description of the ensemble study will be published
elsewhere. Here we focus on one device which showed intriguing and quantitatively informative Kondo behavior. None of the conclusions below depend on whether this particular tube is filled; moreover, the ensemble study and another recent work~\cite{nygard:peapods} strongly suggest that the intercalated C$_{60}$ do not perturb the low-energy transport properties of nanotubes.

\begin{figure}[htbp!]
\includegraphics[0, 0][86mm,70.4mm]{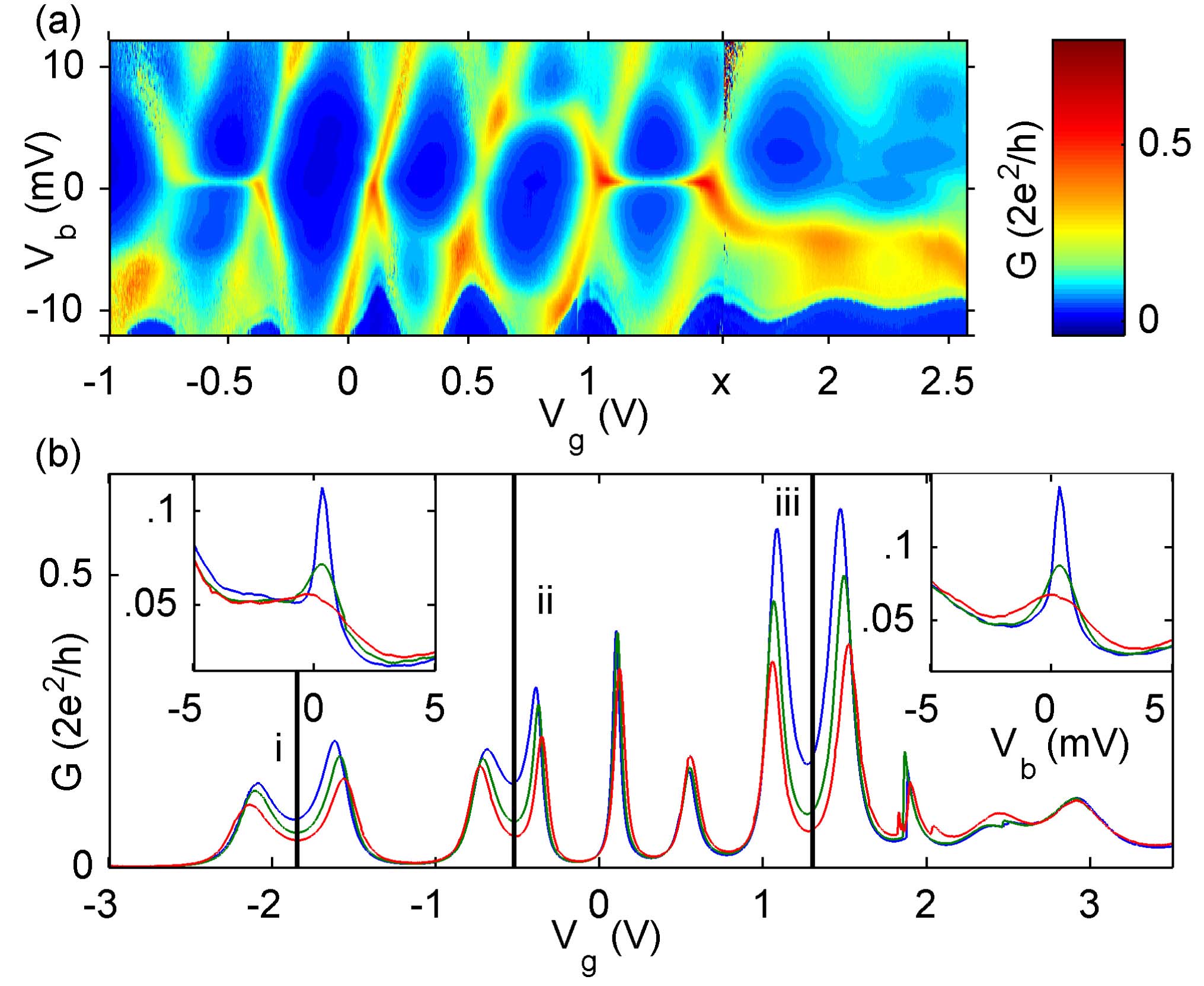}
\caption{\label{f2} (a) Differential conductance $G$ as a function
of bias voltage $V_{\rm b}$, showing zero bias features for odd
electron occupancy. Data to the right of the location marked `x'
have been shifted to account for a random charging event. (b) Linear
conductance as a function of voltage $V_{\rm g}$ on a back gate, at
$T \approx$ 317mK (blue), 1.8K (green) and 6K (red). (Insets) Zero
bias features in valleys ii and iii at the same temperatures.}
\end{figure}

In Coulomb diamond measurements, striking horizontal features --- peaks in differential conductance $G$ at bias voltage $V_{\rm b}=0$, independent of voltage $V_{\rm g}$ on a back gate -- signal the
presence of the Kondo effect (Figure 1a)~\cite{nygard}. These zero-bias anomalies are more prominent at lower temperatures (Figure 1b, insets), as expected, and from their widths we estimate the Kondo temperature to be 1.7K, 1.8K and 1.5K in the middle of valleys (i), (ii) and (iii)~\cite{dgg:prl,nrgdiscussion}.

\begin{figure}[htbp!]
\includegraphics[0,0][86mm,75.7mm]{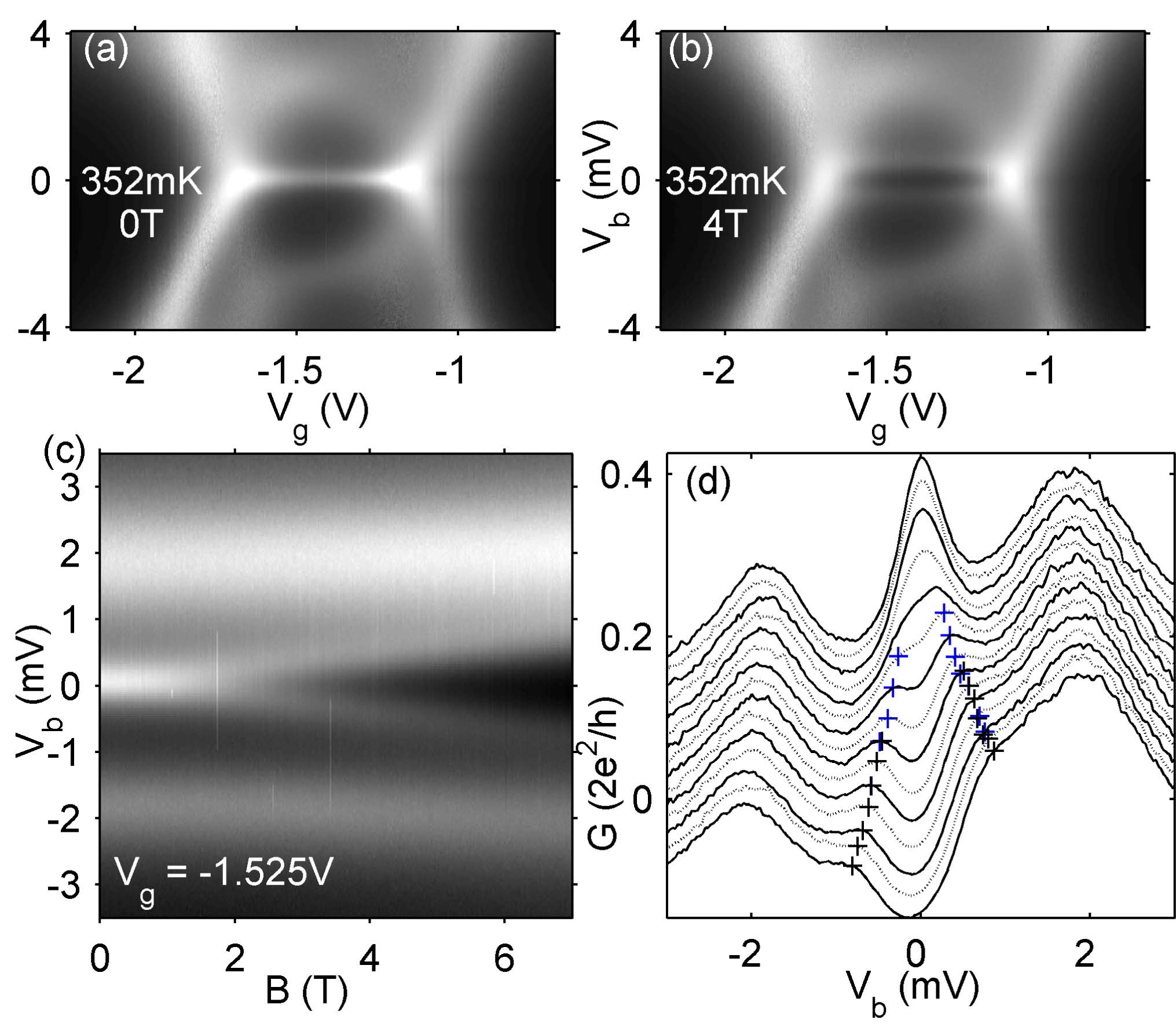}
\caption{\label{f3} (a) Kondo diamond at zero magnetic field. T $\approx$ 352mK. Black is low conductance and white high. (b) The same diamond at 4T. (c) Evolution of the features in the middle of the valley in 2a-b with magnetic field. (d) Slices of the data in 2a in 0.5T steps. (Our data are twenty times denser.) Successive curves are offset vertically by $0.02\times2e^2/h$. Peak locations for the central Kondo peaks were obtained by fitting the data from Figure~2c with two Lorentzians plus a field-independent background (black crosses). A slightly different background is assumed for a similar fit at intermediate field points (blue crosses). Between $4$ and $6T$ the two versions of our fitting procedure produce nearly-indistinguishable results (see also Fig. 3a-d).}
\end{figure}


The Kondo zero-bias conductance peak splits into two peaks upon application of magnetic field (Figure~2), as predicted~\cite{meir} and previously observed~\cite{cronenwett,kogan,cobden}. At high field the spin-resolved peaks in the density of states are predicted to occur at $\pm g\mu_BB \equiv \pm\Delta$~\cite{meir,moore}. At low field the peak energies should be lower~\cite{moore,konik,costi:chap}, with $\delta \rightarrow 2/3 \Delta$ as $B\rightarrow 0$~\cite{logan}. Differing descriptions exist of the crossover between low-field and high-field behavior. Moore and Wen's Bethe-ansatz calculation in an approximate basis converges to the high-field result extremely slowly (logarithmically), predicting for example $\delta \approx 0.9 \Delta$ for $g\mu_B B = 1000 kT_K$~\cite{moore}. Costi, using a density matrix renormalization group approach, predicts a more rapid crossover, with convergence to the high-field result around $g\mu_B B = 20 kT_K$~\cite{costi:chap}. We find precisely linear behavior $2\delta [\mu V] = 240 B [T] - 5 \pm 14$, corresponding to $g = 2.07$, at moderately high fields ($B = 4$ to $7$ T, $g mu_B B = 2.9$ to $5.1 kT_K$). The extracted peak energy falls below linear only at $g\mu_B B = 2.8 kT_K$, a much lower-field crossover than predicted by either of these theories~\cite{cotunnelling,theorydetails}.

\begin{figure}[htbp!]
\includegraphics[0,0][86mm,85.6mm]{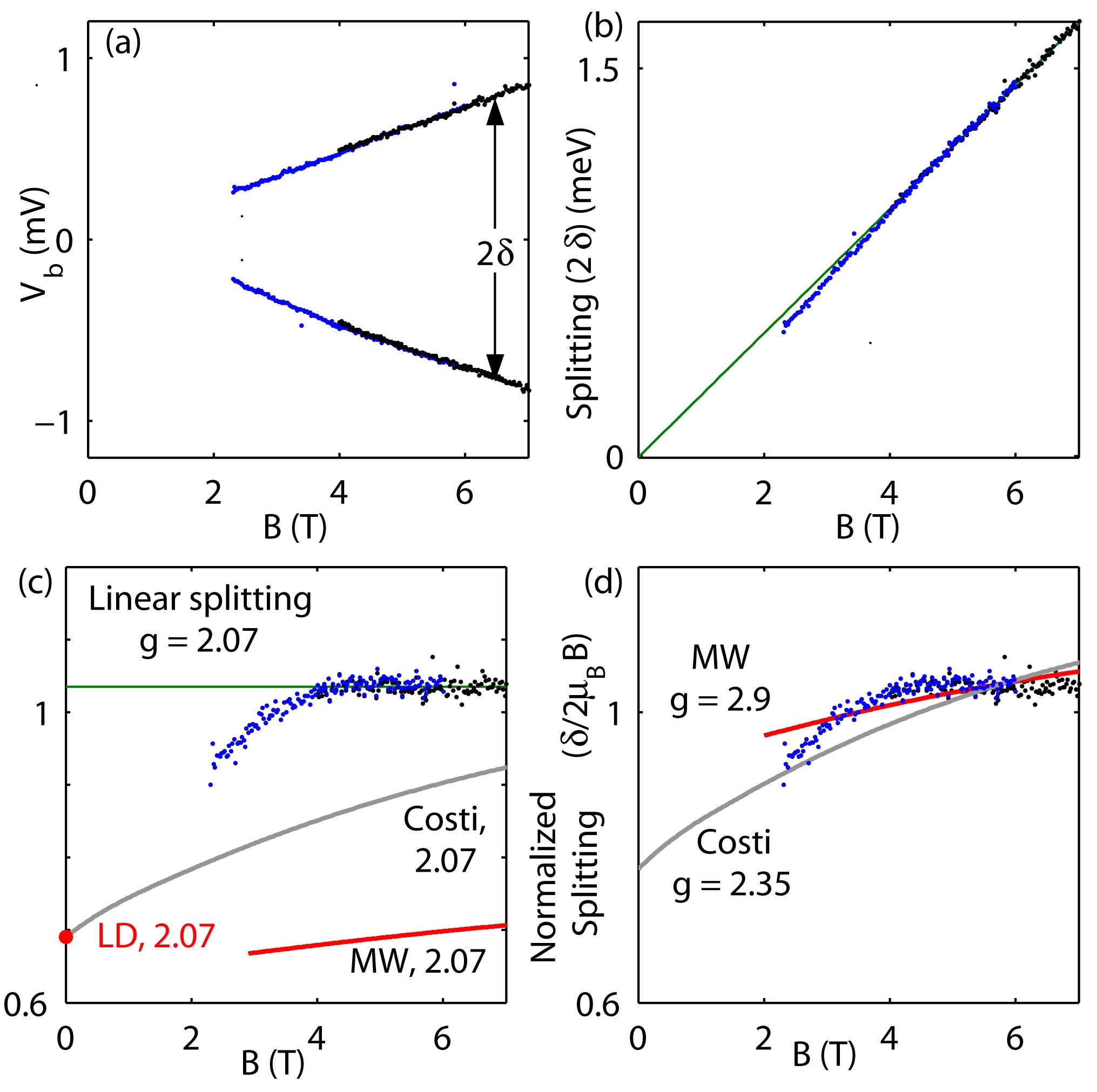}

 \caption{\label{fe}  (a) Peak positions in 2c obtained as described in 2d. Black (blue) dots mark results from the high (intermediate) field fit. (b) Energy difference between peaks from 3a. The green line is a fit to the high field points. 
(c) Splitting from 3a normalized by the naïve Zeeman energy --- $\delta/2\mu_B B$ (black and blue dots). The red line is the Moore-Wen prediction, the grey line is the Costi prediction, the flat green line is the predicted high-field limit, and the red dot marks the Logan-Dickens low-field prediction, all for $g = 2.07$. (d) The data from 3b are reproduced (black and blue dots). The red line is the Moore-Wen prediction and the grey line is the Costi prediction, now using unrealistically large Zeeman coupling $g=2.9$ and 2.35, respectively, to attempt to fit the data.}

\end{figure}

In Figure 3 we compare our data to the Moore-Wen and Costi predictions at both high and low fields. At high fields the data disagree with the theory markedly in the magnitude of the splitting (Figure~3c). If we attempt to correct for this by assuming an unrealistically high g-factor ($g=2.9$ and 2.35 for the respective theories), the theories still predict a more gradual saturation to high field behavior than we observe (Figure~3d).

There are three prominent issues in comparing the various calculations to transport experiments. First, all the theories calculate spin-resolved density of states, whereas experiments measure the sum of the two spin-resolved densities of states. We address this by fitting our measurements
with two Lorentzian peaks and interpreting the extracted peak positions as the positions of the resonances in the spin-resolved density of states. We find that the results are insensitive to details of the fitting procedure for sufficiently high fields, $g\mu_B B > 1.6 k T_K$. At lower fields we do not report peak positions, as fit details are important, and the actual non-Lorentzian (and non-analytic) Kondo lineshape probably influences the extracted peak positions. Lineshape asymmetries at high magnetic field predicted by Rosch \textit{et al}. and Logan-Dickens may also influence our extracted peak positions~\cite{rosch,logan}. Lastly, most theories calculate an equilibrium density of states. In transport experiments a finite bias is applied across the quantum dot, changing the density of states of the system. Konik {\em et al.} extended the Moore-Wen Bethe-ansatz analysis to take into account non-equilibrium effects, and achieved very similar predictions~\cite{konik}. In any case, we expect these effects to be small in our experiments since our dot is coupled much more strongly to one lead than to the other: the ratio of couplings is roughly 8 as judged from our maximum observed conductance.


Since we find a linear splitting at high fields, we assume $\delta = \Delta$ in this range and obtain from the high field data $g = 2.07 \pm 0.02$ (95\% confidence) (Figure 3b)~\cite{seriesresistance}. We believe that ours is the most precise measure to date of the splitting of a Kondo peak with magnetic field in a quantum dot. The value of the g-factor agrees well with previous work on nanotubes~\cite{cobden,tans}. The linearity of the splitting at high field also agrees well with measurements by Cronenwett \textit{et al.}~\cite{cronenwett} in GaAs quantum dots; however, our results differ from those of Kogan \textit{et al.}~\cite{kogan} --- also in GaAs --- who measure $\delta>g\mu_BB$ at all fields extrapolating to a finite zero field value.

Although spin-resolved resonances in principle move away from zero energy even at very low fields, the two resonances initially overlap so that the measured Kondo peak does not visibly split into two until a critical field, $B_c$. Costi has predicted~\cite{costi} that at low temperatures ($T < T_K/2$, satisfied in our case) $g\mu_{B}B_c \approx kT_K$. We observe this splitting first at $B = 2.3T$, modestly higher than the predicted $B_c = 1.5T$, and roughly where we can start to robustly fit peak positions. Previous work in GaAs also found the splitting occurring roughly at $B_c$~\cite{amasha}.

\begin{figure}[htbp] \includegraphics[0,0][86mm,100.5mm]{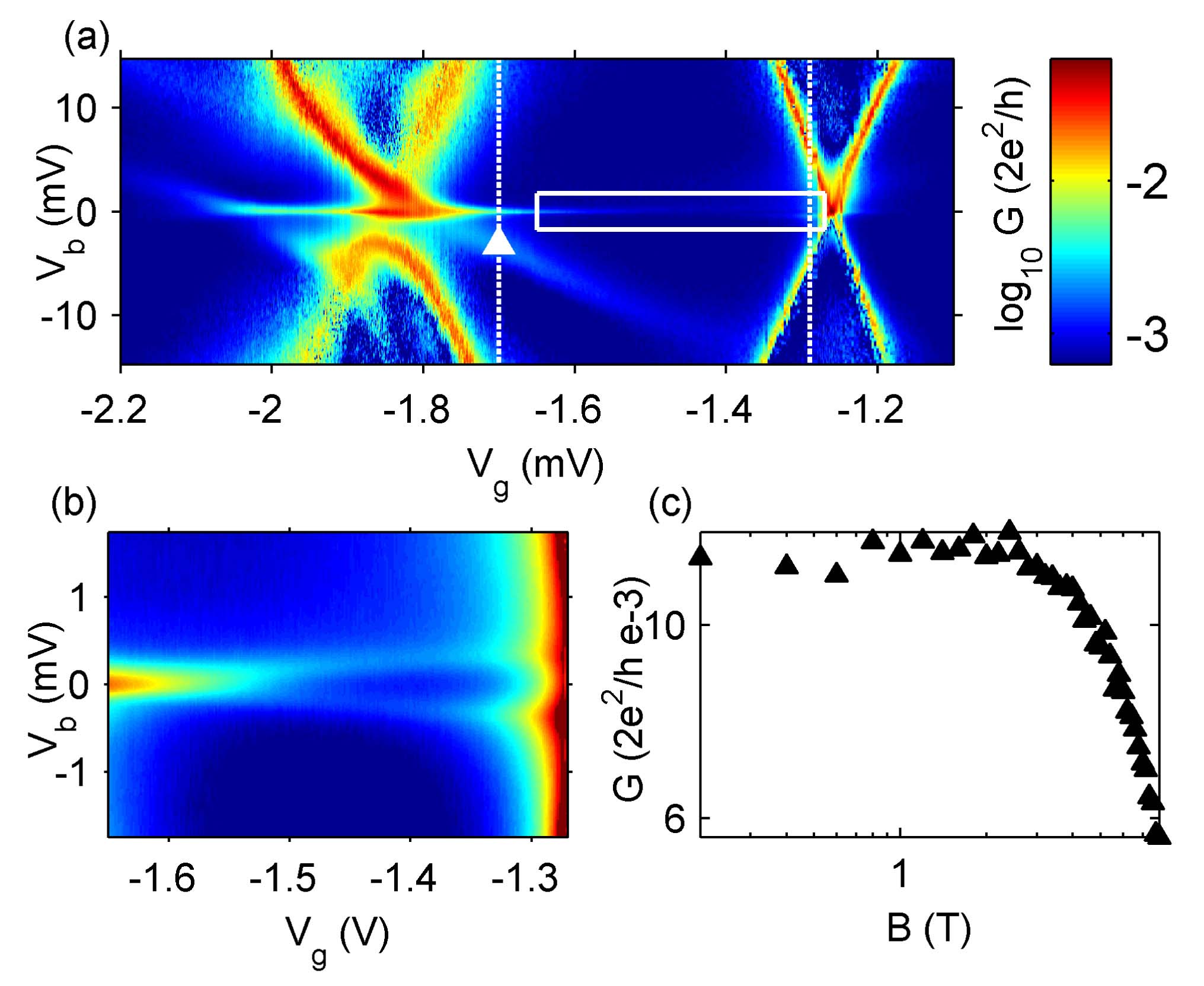} \caption{\label{f4} (a) Zero bias feature in an even Coulomb diamond (parity determined by careful study of ten diamonds on either side).
T $\lesssim$ 250mK. (b) Higher resolution scan of the region in 4a
bounded by the rectangular box. The data indicate a gate-induced
transition between singlet and triplet ground states for the dot.
(c) Magnetic field dependence of the conductance at the point marked
by the white triangle in 4a.} \end{figure}

On a different cooldown of the same device, we observe a Kondo effect for even occupancy of the dot. Enhanced zero-bias conductance on one side of the diamond in question splits into two peaks at
finite bias on the other side (Figure 4a-b) indicating a spin-1 Kondo effect coupled with a gate-induced transition between the triplet and singlet ground states~\cite{kogan:prb}. On the left side of the diamond, the degeneracy of the triplet ground states leads to enhanced conductance at zero bias. On the right, similar to the cotunneling features described above, the conductance is enhanced when the bias voltage coincides with the energy difference between the singlet ground state and the triplet states. Similar gate-induced spin transitions have been previously observed in nanotubes; they may be due to electron-electron correlation effects~\cite{tans:transition}. 


\begin{figure}[htbp!]
\includegraphics[0, 0][86mm,100.5mm]{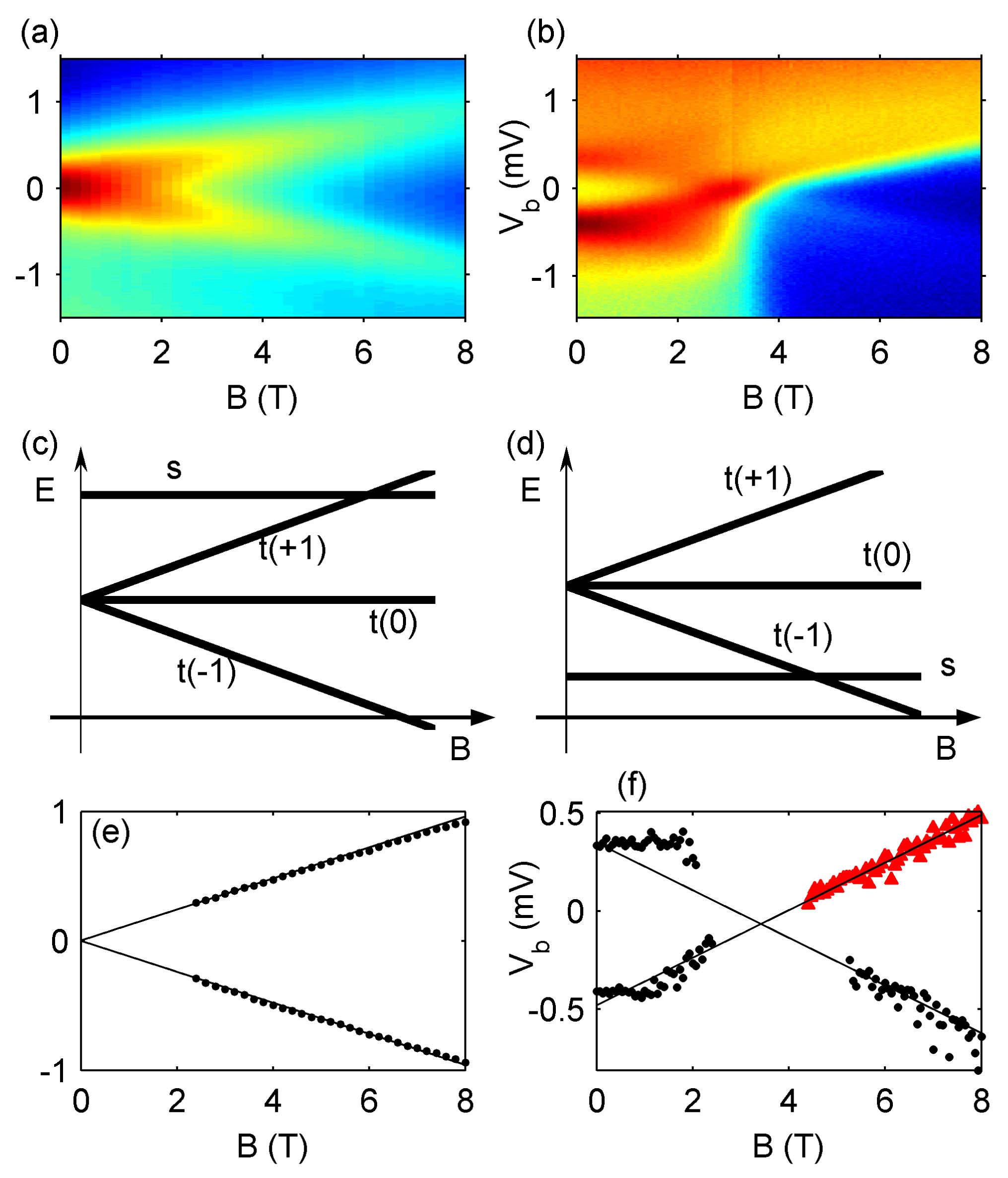}
\caption{\label{f5} (a,b) Magnetic field evolution of conductance versus bias voltage on the left (triplet) and right (singlet) sides of the diamond in 4a. The gate voltages chosen have been marked with dashed lines in 4a. (c,d) Schematic of singlet (s) and triplet (t) energy levels as a function of magnetic field in the configurations corresponding to 5a and 5b repectively. (e) Peak positions in 5a obtained by fitting two Lorentzians. The lines are $\pm g\mu_BB$, with g = 2.07. (f) Analysis of data in 5b. The black circles are peaks obtained using a simple peak-finding function. The red triangles are the locations of steepest slope for the step. The lines are guides to the eye and have slope $\pm g\mu_BB$, with g = 2.07.}
\end{figure}

On the triplet (left) side of the diamond, a magnetic field splits
the peak at zero bias into two peaks at a finite bias corresponding
to the energy difference between the two lowest triplet states
(Figure 5a, c). As with the spin-1/2 Kondo peak, the splitting here
first occurs at slightly larger-than-expected field $B = 1.8$T $>
B_c = 1.1$T~\cite{costi}. There is no explicit theory for the
splitting of the spin-1 Kondo peak, so we determine $T_K$ and $B_c$
in the same way as for spin-1/2.

On the singlet (right) side, two peaks occur at the energy
difference between the singlet and the lowest triplet state (Figure
5b, d). These features can be largely understood by assuming a
linear Zeeman splitting between the various energy levels as
illustrated in Figures 5c and d; however, the transformation in
Figure 5b of the upper peak into a step after the crossing point,
and the insensitivity of both peaks to fields less than 2T, remain
puzzling. We observe several additional strongly gate-sensitive
features in our diamond plots (Figure 4a). At least one of them
appears to be Kondo-related --- the conductance decays
logarithmically with B at high field, and saturates to a constant
value at low field (Figure 4c).

In conclusion, we have observed spin-1 and spin-1/2 Kondo effects in
a carbon nanotube device. We carefully track a spin-1/2 Kondo
zero-bias peak and find that the energy of its spin-resolved
constituents is linear with field at high field, and sublinear at
low field, qualitatively matching predictions. However, the
crossover from sublinear to linear splitting is sharper than
predicted by any existing theory, and occurs at lower-than-expected
field. We also demonstrate the independent gate and magnetic field
tuning of a spin-1 Kondo effect, and find good quantitative
agreement with predictions using the g-factor previously extracted.

We thank S. Amasha, T. Costi, A. Kogan, D. Logan, Y. Meir, J. Moore, and N. Wingreen for helpful conversations. This work was supported by U.S. Air Force Grants No. FA9550-04-1-0384 and F49620-02-1-0383, and a Grant-in-Aid for Scientific Research by the MEXT of Japan. It was performed in part at the Stanford Nanofabrication Facility of NNIN supported by the National Science Foundation under Grant ECS-9731293. CQHL acknowledges support from Gabilan Stanford Graduate and Harvey Fellowships, and DGG fellowships from the Packard and Sloan Foundations.


\begin{thebibliography}{28}
\bibitem[{fab()}]{fabrication}
\bibinfo{note}{Peapods are synthesized by a sublimation
  technique~\cite{kataura}. They are then dispersed by sonication in chloroform
  or ortho-dicholorobenzene and deposited on a highly-doped Si substrate, which
  acts as the gate electrode and has a 500nm thick surface oxide layer. We
  locate the peapods relative to pre-existing alignment marks using atomic
  force microscopy (AFM), and fabricate Pd contacts, 250nm apart in this case,
  using standard electron-beam lithography techniques. This nanotube was
  $\approx$1nm in diameter according to AFM measurements.}
\bibitem[{cot()}]{cotunnelling}
\bibinfo{note}{Conductance peaks at larger bias, which come from inelastic
  cotunneling processes involving orbital rather than spin transitions, are
  observed within the same Coulomb diamond (Figure 2a-b)~\cite{defranceschi}.
  In our fits to the data in Figure 2, we assume that these cotunneling peaks
  do not change with field. Thus the only fitting parameters at each magnetic
  field are associated with the peaks of interest. Note: the device's
  configuration (and hence $T_K$) changed slightly between earlier measurements
  and this measurement.}
\bibitem[{ser()}]{seriesresistance}
\bibinfo{note}{A series resistance of 650$\Omega$ in the leads was taken into
  account.}
\end{thebibliography}
\end{document}